\documentclass{article}
\usepackage{nips11submit_e,times}

\usepackage{amssymb,amsmath}
\usepackage[dvips]{graphicx}
\usepackage{verbatim}

\DeclareMathOperator{\poly}{poly}
\DeclareMathOperator{\Tr}{tr}

\newtheorem{thm}{Theorem}[section]
\newtheorem{prop}[thm]{Proposition}
\newtheorem{lem}[thm]{Lemma}

\newcommand{\set}[1]{\lbrace #1 \rbrace}

\newcommand{\norm}[1]{\lVert #1 \rVert}
\newcommand{\Norm}[1]{\Bigl\lVert #1 \Bigr\rVert}
\newcommand{\tns}{\otimes}

\newcommand{\Bra}[1]{\bigl( #1 \bigr|}
\newcommand{\Ket}[1]{\bigl| #1 \bigr)}

\newcommand{\CC}{\mathbb{C}}

\newcommand{\EE}{\mathbb{E}}

\newcommand{\calA}{\mathcal{A}}
\newcommand{\calB}{\mathcal{B}}
\newcommand{\calI}{\mathcal{I}}

\newcommand{\calM}{\mathcal{M}}
\newcommand{\calX}{\mathcal{X}}
\newcommand{\calY}{\mathcal{Y}}

\newcommand{\Mhat}{\hat{M}}
\newcommand{\vareps}{\varepsilon}

\nipsfinalcopy

\begin{document}

\title{Universal low-rank matrix recovery\\ from Pauli measurements}
\author{Yi-Kai Liu\\
Applied and Computational Mathematics Division\\
National Institute of Standards and Technology\\
Gaithersburg, MD, USA\\
yi-kai.liu@nist.gov}

\date{\today}

\maketitle

\begin{abstract}
We study the problem of reconstructing an unknown matrix $M$ of rank $r$ and dimension $d$ using $O(rd\poly\log d)$ Pauli measurements.  This has applications in quantum state tomography, and is a non-commutative analogue of a well-known problem in compressed sensing:  recovering a sparse vector from a few of its Fourier coefficients.  

We show that almost all sets of $O(rd \log^6 d)$ Pauli measurements satisfy the rank-$r$ restricted isometry property (RIP).  This implies that $M$ can be recovered from a fixed (``universal'') set of Pauli measurements, using nuclear-norm minimization (e.g., the matrix Lasso), with nearly-optimal bounds on the error.  A similar result holds for any class of measurements that use an orthonormal operator basis whose elements have small operator norm.  Our proof uses Dudley's inequality for Gaussian processes, together with bounds on covering numbers obtained via entropy duality.  



\end{abstract}


\section{Introduction}

Low-rank matrix recovery is the following problem:  let $M$ be some unknown matrix of dimension $d$ and rank $r\ll d$, and let $A_1,A_2,\ldots,A_m$ be a set of measurement matrices; then can one reconstruct $M$ from its inner products $\Tr(M^* A_1), \Tr(M^* A_2), \ldots, \Tr(M^* A_m)$?  This problem has many applications in machine learning \cite{Fazel02, Srebro04}, e.g., collaborative filtering (the Netflix problem).  Remarkably, it turns out that for many useful choices of measurement matrices, low-rank matrix recovery is possible, and can even be done efficiently.  For example, when the $A_i$ are Gaussian random matrices, then it is known that $m=O(rd)$ measurements are sufficient to uniquely determine $M$, and furthermore, $M$ can be reconstructed by solving a convex program (minimizing the nuclear norm) \cite{RFP08,FCRP08,CP09}.  Another example is the ``matrix completion'' problem, where the measurements return a random subset of matrix elements of $M$; in this case, $m=O(rd\poly\log d)$ measurements suffice, provided that $M$ satisfies some ``incoherence'' conditions \cite{CandesRecht08, CandesTao09, Gross09, Recht09, NW10}.  

The focus of this paper is on a different class of measurements, known as Pauli measurements.  Here, the $A_i$ are randomly chosen elements of the Pauli basis, a particular orthonormal basis of $\CC^{d\times d}$.  The Pauli basis is a non-commutative analogue of the Fourier basis in $\CC^d$; thus, low-rank matrix recovery using Pauli measurements can be viewed as a generalization of the idea of compressed sensing of sparse vectors using their Fourier coefficients \cite{CandesTao04, RV06}.  In addition, this problem has applications in quantum state tomography, the task of learning an unknown quantum state by performing measurements \cite{Grossetal09}.  This is because most quantum states of physical interest are accurately described by density matrices  that have low rank; and Pauli measurements are especially easy to carry out in an experiment (due to the tensor product structure of the Pauli basis).

In this paper we show stronger results on low-rank matrix recovery from Pauli measurements.  Previously \cite{Grossetal09, Gross09}, it was known that, for every rank-$r$ matrix $M\in\CC^{d\times d}$, almost all choices of $m=O(rd \poly\log d)$ random Pauli measurements will lead to successful recovery of $M$.  Here we show a stronger statement:  there is a fixed (``universal'') set of $m=O(rd \poly\log d)$ Pauli measurements, such that for all rank-$r$ matrices $M\in\CC^{d\times d}$, we have successful recovery.\footnote{Note that in the universal result, $m$ is slightly larger, by a factor of $\poly\log d$.}  We do this by showing that the random Pauli sampling operator obeys the ``restricted isometry property'' (RIP).  Intuitively, RIP says that the sampling operator is an approximate isometry, acting on the set of all low-rank matrices.  In geometric terms, it says that the sampling operator embeds the manifold of low-rank matrices into $O(rd \poly\log d)$ dimensions, with low distortion in the 2-norm.

RIP for low-rank matrices is a very strong property, and prior to this work, it was only known to hold for very unstructured types of random measurements, such as Gaussian measurements \cite{RFP08}, which are unsuitable for most applications.  RIP was known to fail in the matrix completion case, and whether it held for Pauli measurements was an open question.  Once we have established RIP for Pauli measurements, we can use known results \cite{RFP08,FCRP08,CP09} to show low-rank matrix recovery from a universal set of Pauli measurements.  In particular, using \cite{CP09}, we can get nearly-optimal universal bounds on the error of the reconstructed density matrix, when the data are noisy; and we can even get bounds on the recovery of arbitrary (not necessarily low-rank) matrices.  These RIP-based bounds are qualitatively stronger than those obtained using ``dual certificates'' \cite{CandesPlan-old} (though the latter technique is applicable in some situations where RIP fails).

In the context of quantum state tomography, this implies that, given a quantum state that consists of a low-rank component $M_r$ plus a residual full-rank component $M_c$, we can reconstruct $M_r$ up to an error that is not much larger than $M_c$.  In particular, let $\norm{\cdot}_*$ denote the nuclear norm, and let $\norm{\cdot}_F$ denote the Frobenius norm.  Then the error can be bounded in the nuclear norm by $O(\norm{M_c}_*)$ (assuming noiseless data), and it can be bounded in the Frobenius norm by $O(\norm{M_c}_F \poly\log d)$ (which holds even with noisy data\footnote{However, this bound is not universal.}).  This shows that our reconstruction is nearly as good as the best rank-$r$ approximation to $M$ (which is given by the truncated SVD).  In addition, a completely arbitrary quantum state can be reconstructed up to an error of $O(1/\sqrt{r})$ in Frobenius norm.  
Lastly, the RIP gives some insight into the optimal design of tomography experiments, in particular, the tradeoff between the number of measurement settings (which is essentially $m$), and the number of repetitions of the experiment at each setting (which determines the statistical noise that enters the data) \cite{tradeoff11}.

These results can be generalized beyond the class of Pauli measurements.  Essentially, one can replace the Pauli basis with any orthonormal basis of $\CC^{d\times d}$ that is \textit{incoherent}, i.e., whose elements have small operator norm (of order $O(1/\sqrt{d})$, say); a similar generalization was noted in the earlier results of \cite{Gross09}.  Also, our proof shows that the RIP actually holds in a slightly stronger sense:  it holds not just for all rank-$r$ matrices, but for all matrices $X$ that satisfy $\norm{X}_* \leq \sqrt{r}\norm{X}_F$.  


To prove this result, we combine a number of techniques that have appeared elsewhere.  RIP results were previously known for Gaussian measurements and some of their close relatives \cite{RFP08}.  Also, restricted strong convexity (RSC), a similar but somewhat weaker property, was recently shown in the context of the matrix completion problem (with additional ``non-spikiness'' conditions) \cite{NW10}.  These results follow from covering arguments (i.e., using a concentration inequality to upper-bound the failure probability on each individual low-rank matrix $X$, and then taking the union bound over all such $X$).  Showing RIP for Pauli measurements seems to be more delicate, however.  Pauli measurements have more structure and less randomness, so the concentration of measure phenomena are weaker, and the union bound no longer gives the desired result.  

Instead, one must take into account the favorable correlations between the behavior of the sampling operator on different matrices --- intuitively, if two low-rank matrices $M$ and $M'$ have overlapping supports, then good behavior on $M$ is positively correlated with good behavior on $M'$.  This can be done by transforming the problem into a Gaussian process, and using Dudley's entropy bound.  This is the same approach used in classical compressed sensing, to show RIP for Fourier measurements \cite{RV06, CandesTao04}.  The key difference is that in our case, the Gaussian process is indexed by low-rank matrices, rather than sparse vectors.  To bound the correlations in this process, one then needs to bound the covering numbers of the nuclear norm ball (of matrices), rather than the $\ell_1$ ball (of vectors).  This requires a different technique, using entropy duality, which is due to Gu\'edon et al \cite{Guedon08}.   (See also the related work in \cite{Aubrun09}.)  

As a side note, we remark that matrix recovery can sometimes fail because there exist large sets of up to $d$ Pauli matrices that all commute, i.e., they have a simultaneous eigenbasis $\phi_1,\ldots,\phi_d$.  (These $\phi_i$ are of interest in quantum information --- they are called stabilizer states \cite{NC}.)  If one were to measure such a set of Pauli's, one would gain complete knowledge about the diagonal elements of the unknown matrix $M$ in the $\phi_i$ basis, but one would learn nothing about the off-diagonal elements.  This is reminiscent of the difficulties that arise in matrix completion.  However, in our case, these pathological cases turn out to be rare, since it is unlikely that a random subset of Pauli matrices will all commute.

Finally, we note that there is a large body of related work on estimating a low-rank matrix by solving a regularized convex program; see, e.g., \cite{RT09,
KLT10}.

This paper is organized as follows.  In section 2, we state our results precisely, and discuss some specific applications to quantum state tomography.  In section 3 we prove the RIP for Pauli matrices, and in section 4 we discuss some directions for future work.  
Some technical details appear in sections \ref{app3} 
and \ref{sec4}. 

\textbf{Notation:}  For vectors, $\norm{\cdot}_2$ denotes the $\ell_2$ norm.  For matrices, $\norm{\cdot}_p$ denotes the Schatten $p$-norm, $\norm{X}_p = (\sum_i \sigma_i(X)^p)^{1/p}$, where $\sigma_i(X)$ are the singular values of $X$.  In particular, $\norm{\cdot}_* = \norm{\cdot}_1$ is the trace or nuclear norm, $\norm{\cdot}_F = \norm{\cdot}_2$ is the Frobenius norm, and $\norm{\cdot} = \norm{\cdot}_\infty$ is the operator norm.  Finally, for matrices, $A^*$ is the adjoint of $A$, and $(\cdot,\cdot)$ is the Hilbert-Schmidt inner product, $(A,B) = \Tr(A^* B)$.  Calligraphic letters denote superoperators acting on matrices.  Also, $\Ket{A}\Bra{A}$ is the superoperator that maps every matrix $X\in\CC^{d\times d}$ to the matrix $A\Tr(A^*X)$.


\section{Our Results}

We will consider the following approach to low-rank matrix recovery.  Let $M\in\CC^{d\times d}$ be an unknown matrix of rank at most $r$.  Let $W_1,\ldots,W_{d^2}$ be an orthonormal basis for $\CC^{d\times d}$, with respect to the inner product $(A,B) = \Tr(A^* B)$.  We choose $m$ basis elements, $S_1,\ldots,S_m$, iid uniformly at random from $\set{W_1,\ldots,W_{d^2}}$ (``sampling with replacement'').  We then observe the coefficients $(S_i,M)$.  From this data, we want to reconstruct $M$.

For this to be possible, the measurement matrices $W_i$ must be ``incoherent'' with respect to $M$.  Roughly speaking, this means that the inner products $(W_i,M)$ must be small.  Formally, we say that the basis $W_1,\ldots,W_{d^2}$ is \textit{incoherent} if the $W_i$ all have small operator norm, 
\begin{equation}
\label{eqn-smallopnorm}
\norm{W_i} \leq K/\sqrt{d}, 
\end{equation}
where $K$ is a constant.\footnote{Note that $\norm{W_i}$ is the maximum inner product between $W_i$ and any rank-1 matrix $M$ (normalized so that $\norm{M}_F = 1$).}  (This assumption was also used in \cite{Gross09}.)  

Before proceeding further, let us sketch the connection between this problem and quantum state tomography.  Consider a system of $n$ qubits, with Hilbert space dimension $d = 2^n$.  We want to learn the state of the system, which is described by a density matrix $\rho\in\CC^{d\times d}$; $\rho$ is positive semidefinite, has trace 1, and has rank $r\ll d$ when the state is nearly pure.  There is a class of convenient (and experimentally feasible) measurements, which are described by Pauli matrices (also called Pauli observables).  These are matrices of the form $P_1\tns\cdots\tns P_n$, where $\tns$ denotes the tensor product (Kronecker product), and each $P_i$ is a $2\times 2$ matrix chosen from the following four possibilities:  
\begin{equation}
I = \begin{pmatrix} 1&0\\ 0&1 \end{pmatrix}, \quad
\sigma_x = \begin{pmatrix} 0&1\\ 1&0 \end{pmatrix}, \quad
\sigma_y = \begin{pmatrix} 0&-i\\ i&0 \end{pmatrix}, \quad
\sigma_z = \begin{pmatrix} 1&0\\ 0&-1 \end{pmatrix}.
\end{equation}
One can estimate expectation values of Pauli observables, which are given by $(\rho,(P_1\tns\cdots\tns P_n))$.  This is a special case of the above measurement model, where the measurement matrices $W_i$ are the (scaled) Pauli observables $(P_1\tns\cdots\tns P_n)/\sqrt{d}$, and they are incoherent with $\norm{W_i} \leq K/\sqrt{d}$, $K=1$.

Now we return to our discussion of the general problem.  We choose $S_1,\ldots,S_m$ iid uniformly at random from $\set{W_1,\ldots,W_{d^2}}$, and we define the \textit{sampling operator} $\calA:\: \CC^{d\times d} \rightarrow \CC^m$ as 
\begin{equation}
\label{eqn-A}
(\calA(X))_i = \tfrac{d}{\sqrt{m}} \Tr(S_i^* X), \quad i=1,\ldots,m.
\end{equation}
The normalization is chosen so that $\EE \calA^*\calA = \calI$.  (Note that $\calA^*\calA = \sum_{j=1}^m \Ket{S_j}\Bra{S_j} \cdot \tfrac{d^2}{m}$.)  

We assume we are given the data 
$y = \calA(M)+z$, 
where $z\in\CC^m$ is some (unknown) noise contribution.  We will construct an estimator $\Mhat$ by minimizing the nuclear norm, subject to the constraints specified by $y$.  (Note that one can view the nuclear norm as a convex relaxation of the rank function --- thus these estimators can be computed efficiently.)  One approach is the matrix Dantzig selector:
\begin{equation}
\label{eqn-tracemin}
\Mhat = \arg\min_X \: \norm{X}_* \text{ such that } \norm{\calA^*(y-\calA(X))} \leq \lambda.
\end{equation}
Alternatively, one can solve a regularized least-squares problem, also called the matrix Lasso:
\begin{equation}
\label{eqn-lasso}
\Mhat = \arg\min_X \tfrac{1}{2} \norm{\calA(X)-y}_2^2 + \mu\norm{X}_*.
\end{equation}
Here, the parameters $\lambda$ and $\mu$ are set according to the strength of the noise component $z$ (we will discuss this later).  We will be interested in bounding the error of these estimators.  To do this, we will show that the sampling operator $\calA$ satisfies the restricted isometry property (RIP).

\subsection{RIP for Pauli Measurements}

Fix some constant $0 \leq \delta < 1$.  Fix $d$, and some set $U \subset \CC^{d\times d}$.  We say that $\calA$ satisfies the \textit{restricted isometry property} (RIP) over $U$ if, for all $X \in U$, we have 
\begin{equation}
\label{eqn-RIP}
(1-\delta) \norm{X}_F \leq \norm{\calA(X)}_2 \leq (1+\delta) \norm{X}_F.
\end{equation}
(Here, $\norm{\calA(X)}_2$ denotes the $\ell_2$ norm of a vector, while $\norm{X}_F$ denotes the Frobenius norm of a matrix.)  When $U$ is the set of all $X \in \CC^{d\times d}$ with rank $r$, this is precisely the notion of RIP studied in \cite{RFP08, CP09}.  We will show that Pauli measurements satisfy the RIP over a slightly larger set (the set of all $X\in\CC^{d\times d}$ such that $\norm{X}_* \leq \sqrt{r} \norm{X}_F$), provided the number of measurements $m$ is at least $\Omega(rd\poly\log d)$.  This result generalizes to measurements in any basis with small operator norm.
\begin{thm}
\label{thm-rip}
Fix some constant $0 \leq \delta < 1$.  Let $\set{W_1,\ldots,W_{d^2}}$ be an orthonormal basis for $\CC^{d\times d}$ that is incoherent in the sense of (\ref{eqn-smallopnorm}).  Let $m = CK^2\cdot rd \log^6 d$, for some constant $C$ that depends only on $\delta$, $C = O(1/\delta^2)$.  Let $\calA$ be defined as in (\ref{eqn-A}).  Then, with high probability (over the choice of $S_1,\ldots,S_m$), $\calA$ satisfies the RIP over the set of all $X\in\CC^{d\times d}$ such that $\norm{X}_* \leq \sqrt{r} \norm{X}_F$.  Furthermore, the failure probability is exponentially small in $\delta^2 C$.  
\end{thm}

We will prove this theorem in section 3.  In the remainder of this section, we discuss its applications to low-rank matrix recovery, and quantum state tomography in particular.

\subsection{Applications}

By combining Theorem \ref{thm-rip} with previous results \cite{RFP08, FCRP08, CP09}, we immediately obtain bounds on the accuracy of the matrix Dantzig selector (\ref{eqn-tracemin}) and the matrix Lasso (\ref{eqn-lasso}).  In particular, for the first time we can show \textit{universal} recovery of low-rank matrices via Pauli measurements, and near-optimal bounds on the accuracy of the reconstruction when the data is noisy \cite{CP09}.  (Similar results hold for measurements in any incoherent operator basis.)  These RIP-based results improve on the earlier results based on dual certificates \cite{Grossetal09, Gross09, CandesPlan-old}.  See \cite{RFP08, FCRP08, CP09} for details.  

Here, we will sketch a couple of these results that are of particular interest for quantum state tomography.  Here, $M$ is the density matrix describing the state of a quantum mechanical object, and $\calA(M)$ is a vector of Pauli expectation values for the state $M$.  ($M$ has some additional properties:  it is positive semidefinite, and has trace 1; thus $\calA(M)$ is a real vector.)  There are two main issues that arise.  First, $M$ is not precisely low-rank.  In many situations, the \textit{ideal} state has low rank (for instance, a pure state has rank 1); however, for the \textit{actual} state observed in an experiment, the density matrix $M$ is full-rank with decaying eigenvalues.  Typically, we will be interested in obtaining a good low-rank approximation to $M$, ignoring the tail of the spectrum.

Secondly, the measurements of $\calA(M)$ are inherently noisy.  We do not observe $\calA(M)$ directly; rather, we estimate each entry $(\calA(M))_i$ by preparing many copies of the state $M$, measuring the Pauli observable $S_i$ on each copy, and averaging the results.  Thus, we observe $y_i = (\calA(M))_i + z_i$, where $z_i$ is binomially distributed.  When the number of experiments being averaged is large, $z_i$ can be approximated by Gaussian noise.  We will be interested in getting an estimate of $M$ that is stable with respect to this noise.  (We remark that one can also reduce the statistical noise by performing more repetitions of each experiment.  This suggests the possibility of a tradeoff between the accuracy of estimating each parameter, and the number of parameters one chooses to measure overall.  This will be discussed elsewhere \cite{tradeoff11}.)

We would like to reconstruct $M$ up to a small error in the nuclear or Frobenius norm.  Let $\Mhat$ be our estimate.  Bounding the error in nuclear norm implies that, for any measurement allowed by quantum mechanics, the probability of distinguishing the state $\Mhat$ from $M$ is small.  Bounding the error in Frobenius norm implies that the difference $\Mhat-M$ is highly ``mixed'' (and thus does not contribute to the coherent or ``quantum'' behavior of the system).

We now sketch a few results from \cite{FCRP08,CP09} that apply to this situation.  Write $M = M_r+M_c$, where $M_r$ is a rank-$r$ approximation to $M$, corresponding to the $r$ largest singular values of $M$, and $M_c$ is the residual part of $M$ (the ``tail'' of $M$).  Ideally, our goal is to estimate $M$ up to an error that is not much larger than $M_c$.  First, we can bound the error in nuclear norm (assuming the data has no noise):
\begin{prop}
\label{prop-errorbound-nuc}
(Theorem 5 from \cite{FCRP08}) 
Let $\calA:\: \CC^{d\times d} \rightarrow \CC^m$ be the random Pauli sampling operator, with $m = C rd \log^6 d$, for some absolute constant $C$.  Then, with high probability over the choice of $\calA$, the following holds:

Let $M$ be any matrix in $\CC^{d\times d}$, and write $M = M_r + M_c$, as described above.  
Say we observe $y = \calA(M)$, with no noise.  Let $\Mhat$ be the Dantzig selector (\ref{eqn-tracemin}) with $\lambda = 0$.  Then 
\begin{equation}
\label{eqn-errorbound-nuc}
\norm{\Mhat-M}_* \leq C'_0 \norm{M_c}_*, 
\end{equation}
where $C'_0$ is an absolute constant.
\end{prop}

We can also bound the error in Frobenius norm, allowing for noisy data:
\begin{prop}
\label{prop-errorbound}
(Lemma 3.2 from \cite{CP09}) 
Assume the same set-up as above, but say we observe $y = \calA(M)+z$, where $z \sim N(0,\sigma^2 I)$.  Let $\Mhat$ be the Dantzig selector (\ref{eqn-tracemin}) with $\lambda = 8\sqrt{d}\sigma$, or the Lasso (\ref{eqn-lasso}) with $\mu = 16\sqrt{d}\sigma$.  Then, with high probability over the noise $z$, 
\begin{equation}
\label{eqn-errorbound}
\norm{\Mhat-M}_F \leq C_0 \sqrt{rd} \sigma + C_1 \norm{M_c}_* / \sqrt{r}, 
\end{equation}
where $C_0$ and $C_1$ are absolute constants.
\end{prop}
This bounds the error of $\Mhat$ in terms of the noise strength $\sigma$ and the size of the tail $M_c$.  It is universal:  one sampling operator $\calA$ works for all matrices $M$.  While this bound may seem unnatural because it mixes different norms, it can be quite useful.  
When $M$ actually is low-rank (with rank $r$), then $M_c = 0$, and the bound (\ref{eqn-errorbound}) becomes particularly simple.  The dependence on the noise strength $\sigma$ is known to be nearly minimax-optimal \cite{CP09}.  Furthermore, when some of the singular values of $M$ fall below the ``noise level'' $\sqrt{d}\sigma$, one can show a tighter bound, with a nearly-optimal bias-variance tradeoff; see Theorem 2.7 in \cite{CP09} for details.  

On the other hand, when $M$ is full-rank, then the error of $\Mhat$ depends on the behavior of the tail $M_c$.  We will consider a couple of cases.  First, suppose we do not assume anything about $M$, besides the fact that it is a density matrix for a quantum state.  Then $\norm{M}_* = 1$, hence $\norm{M_c}_* \leq 1 - \tfrac{r}{d}$, and we can use (\ref{eqn-errorbound}) to get 
$\norm{\Mhat-M}_F \leq C_0 \sqrt{rd} \sigma + \frac{C_1}{\sqrt{r}}$.
Thus, even for \textit{arbitrary} (not necessarily low-rank) quantum states, the estimator $\Mhat$ gives nontrivial results.  
The $O(1/\sqrt{r})$ term can be interpreted as the penalty for only measuring an incomplete subset of the Pauli observables.

Finally, consider the case where $M$ is full-rank, but we do know that the tail $M_c$ is small.  If we know that $M_c$ is small in nuclear norm, then we can use equation (\ref{eqn-errorbound}).  However, if we know that $M_c$ is small in Frobenius norm, one can give a different bound, using ideas from \cite{CP09}, as follows.  

\begin{prop}
\label{prop-errorbound2}
Let $M$ be any matrix in $\CC^{d\times d}$, with singular values $\sigma_1(M) \geq \cdots \geq \sigma_d(M)$.  

Choose a random Pauli sampling operator $\calA:\: \CC^{d\times d} \rightarrow \CC^m$, with $m = C rd \log^6 d$, for some absolute constant $C$.  Say we observe $y = \calA(M)+z$, where $z \sim N(0,\sigma^2 I)$.  Let $\Mhat$ be the Dantzig selector (\ref{eqn-tracemin}) with $\lambda = 16\sqrt{d}\sigma$, or the Lasso (\ref{eqn-lasso}) with $\mu = 32\sqrt{d}\sigma$.  Then, with high probability over the choice of $\calA$ and the noise $z$, 
\begin{equation}
\label{eqn-errorbound2}
\norm{\Mhat-M}_F^2 \leq C_0 \sum_{i=1}^r \min(\sigma_i^2(M), d\sigma^2) + C_2(\log^6 d) \sum_{i=r+1}^d \sigma_i^2(M), 
\end{equation}
where $C_0$ and $C_2$ are absolute constants.
\end{prop}

This bound can be interpreted as follows.  
The first term expresses the bias-variance tradeoff for estimating $M_r$, while the second term depends on the Frobenius norm of $M_c$.  (Note that the $\log^6 d$ factor may not be tight.)  
In particular, this implies:
$\norm{\Mhat-M}_F \leq \sqrt{C_0} \sqrt{rd}\sigma + \sqrt{C_2}(\log^3 d) \norm{M_c}_F$.
This can be compared with equation (\ref{eqn-errorbound}) (involving $\norm{M_c}_*$).  
This bound will be better when $\norm{M_c}_F \ll \norm{M_c}_*$, i.e., when the tail $M_c$ has slowly-decaying eigenvalues (in physical terms, it is highly mixed).  

Proposition \ref{prop-errorbound2} is an adaptation of Theorem 2.8 in \cite{CP09}.  We sketch the proof in section \ref{sec4}. 
Note that this bound is not universal:  it shows that for all matrices $M$, a random choice of the sampling operator $\calA$ is likely to work.


\section{Proof of the RIP for Pauli Measurements}

We now prove Theorem \ref{thm-rip}.  The general approach involving Dudley's entropy bound is similar to \cite{RV06}, while the technical part of the proof (bounding certain covering numbers) uses ideas from \cite{Guedon08}.  We summarize the argument here; the details are given in section \ref{app3}. 

\subsection{Overview}

Let 
$U_2 = \set{X\in\CC^{d\times d} \;|\; \norm{X}_F \leq 1, \; \norm{X}_* \leq \sqrt{r}\norm{X}_F}$.
Let $\calB$ be the set of all self-adjoint linear operators from $\CC^{d\times d}$ to $\CC^{d\times d}$, and define the following norm on $\calB$:
\begin{equation}
\norm{\calM}_{(r)} = \sup_{X\in U_2} |(X,\calM X)|.
\end{equation}
(Suppose $r\geq 2$, which is sufficient for our purposes.  It is straightforward to show that $\norm{\cdot}_{(r)}$ is a norm, and that $\calB$ is a Banach space with respect to this norm.)  Then let us define 
\begin{equation}
\vareps_r(\calA) = \norm{\calA^*\calA-\calI}_{(r)}.  
\end{equation}
By an elementary argument, in order to prove RIP, it suffices to show that $\vareps_r(\calA) < 2\delta - \delta^2$.  We will proceed as follows:  we will first bound $\EE \vareps_r(\calA)$, then show that $\vareps_r(\calA)$ is concentrated around its mean.

Using a standard symmetrization argument, we have that 
$\EE \vareps_r(\calA) \leq 2\EE \Norm{\sum_{j=1}^m \vareps_j \Ket{S_j}\Bra{S_j} \tfrac{d^2}{m}}_{(r)}$,
where the $\vareps_j$ are Rademacher (iid $\pm 1$) random variables.  Here the round ket notation $\Ket{S_j}$ means we view the matrix $S_j$ as an element of the vector space $\CC^{d^2}$ with Hilbert-Schmidt inner product; the round bra $\Bra{S_j}$ denotes the adjoint element in the (dual) vector space.  

Now we use the following lemma, which we will prove later.  This bounds the expected magnitude in $(r)$-norm of a Rademacher sum of a fixed collection of operators $V_1,\ldots,V_m$ that have small operator norm.
\begin{lem}
\label{lem-key}
Let $m\leq d^2$.  Fix some $V_1,\ldots,V_m \in \CC^{d\times d}$ that have uniformly bounded operator norm, $\norm{V_i} \leq K$ (for all $i$).  Let $\vareps_1,\ldots,\vareps_m$ be iid uniform $\pm 1$ random variables.  Then 
\begin{equation}
\EE_\vareps \Norm{\sum_{i=1}^m \vareps_i\Ket{V_i}\Bra{V_i}}_{(r)}
 \leq C_5\cdot \Norm{\sum_{i=1}^m \Ket{V_i}\Bra{V_i}}_{(r)}^{1/2}, 
\end{equation}
where $C_5 = \sqrt{r}\cdot C_4 K \log^{5/2} d \log^{1/2} m$ and $C_4$ is some universal constant.
\end{lem}

After some algebra, one gets that 
$\EE\vareps_r(\calA)
 \leq 2 (\EE\vareps_r(\calA) + 1)^{1/2} \cdot C_5 \cdot \sqrt{\tfrac{d}{m}}$, 
where $C_5 = \sqrt{r}\cdot C_4 K \log^3 d$.  By finding the roots of this quadratic equation, we get the following bound on $\EE\vareps_r(\calA)$.  
Let $\lambda \geq 1$.  Assume that 
$m \geq \lambda d (2C_5)^2 = \lambda \cdot 4C_4^2 \cdot dr \cdot K^2 \log^6 d$.  
Then we have the desired result:
\begin{equation}
\EE\vareps_r(\calA) \leq \tfrac{1}{\lambda} + \tfrac{1}{\sqrt{\lambda}}.
\end{equation}

It remains to show that $\vareps_r(\calA)$ is concentrated around its expectation.  For this we use a concentration inequality from \cite{LT91} for sums of independent symmetric random variables that take values in some Banach space.  See section \ref{app3} 
for details.


\subsection{Proof of Lemma \ref{lem-key} (bounding a Rademacher sum in $(r)$-norm)}

Let $L_0 = \EE_\vareps \norm{\sum_{i=1}^m \vareps_i\Ket{V_i}\Bra{V_i}}_{(r)}$; this is the quantity we want to bound.  Using a standard comparison principle, we can replace the $\pm 1$ random variables $\vareps_i$ with iid $N(0,1)$ Gaussian random variables $g_i$; then we get 
\begin{equation}
L_0 \leq \EE_g \sup_{X\in U_2} \sqrt{\tfrac{\pi}{2}} |G(X)|, \quad
G(X) = \sum_{i=1}^m g_i |(V_i,X)|^2.
\end{equation}
The random variables $G(X)$ (indexed by $X\in U_2$) form a Gaussian process, and $L_0$ is upper-bounded by the expected supremum of this process.  Using the fact that $G(0)=0$ and $G(\cdot)$ is symmetric, and Dudley's inequality (Theorem 11.17 in \cite{LT91}), we have 
\begin{equation}
L_0 \leq \sqrt{2\pi} \EE_g \sup_{X\in U_2} G(X)
    \leq 24 \sqrt{2\pi} \int_0^\infty \log^{1/2} N(U_2,d_G,\vareps) d\vareps, 
\end{equation}
where $N(U_2,d_G,\vareps)$ is a covering number (the number of balls in $\CC^{d\times d}$ of radius $\vareps$ in the metric $d_G$ that are needed to cover the set $U_2$), and the metric $d_G$ is given by 
\begin{equation}
d_G(X,Y) = \Bigl( \EE[(G(X)-G(Y))^2] \Bigr)^{1/2}.
\end{equation}

Define a new norm (actually a semi-norm) $\norm{\cdot}_X$ on $\CC^{d\times d}$, as follows:
\begin{equation}
\norm{M}_X = \max_{i=1,\ldots,m} |(V_i,M)|.
\end{equation}
We use this to upper-bound the metric $d_G$.  An elementary calculation shows that $d_G(X,Y) \leq 2R\norm{X-Y}_X$, where $R = \norm{\sum_{i=1}^m \Ket{V_i}\Bra{V_i}}_{(r)}^{1/2}$.  This lets us upper-bound the covering numbers in $d_G$ with covering numbers in $\norm{\cdot}_X$:
\begin{equation}
N(U_2, d_G, \vareps) \leq N(U_2, \norm{\cdot}_X, \tfrac{\vareps}{2R}) = N(\tfrac{1}{\sqrt{r}}U_2, \norm{\cdot}_X, \tfrac{\vareps}{2R\sqrt{r}}).
\end{equation}

We will now bound these covering numbers.  First, we introduce some notation:  let $\norm{\cdot}_p$ denote the Schatten $p$-norm on $\CC^{d\times d}$, and let $B_p$ be the unit ball in this norm.  Also, let $B_X$ be the unit ball in the $\norm{\cdot}_X$ norm.  

Observe that 
$\tfrac{1}{\sqrt{r}}U_2 \subseteq B_1 \subseteq K\cdot B_X$.
(The second inclusion follows because $\norm{M}_X \leq \max_{i=1,\ldots,m} \norm{V_i} \norm{M}_* \leq K \norm{M}_*$.)  This gives a simple bound on the covering numbers:
\begin{equation}
N(\tfrac{1}{\sqrt{r}}U_2, \norm{\cdot}_X, \vareps) \leq N(B_1, \norm{\cdot}_X, \vareps)
 \leq N(K\cdot B_X, \norm{\cdot}_X, \vareps).
\end{equation}
This is 1 when $\vareps\geq K$.  So, in Dudley's inequality, 
we can restrict the integral to the interval $[0,K]$.

When $\vareps$ is small, we will use the following simple bound (equation (5.7) in \cite{Pisier}):
\begin{equation}
\label{eqn-coversmall}
N(K\cdot B_X, \norm{\cdot}_X, \vareps) \leq (1+\tfrac{2K}{\vareps})^{2d^2}.
\end{equation}
When $\vareps$ is large, we will use a more sophisticated bound based on Maurey's empirical method and entropy duality, which is due to \cite{Guedon08} (see also \cite{Aubrun09}):
\begin{equation}
\label{eqn-coverlarge}
N(B_1, \norm{\cdot}_X, \vareps) \leq \exp(\tfrac{C_1^2 K^2}{\vareps^2} \log^3 d \log m), \quad \text{for some constant $C_1$}.
\end{equation}
We defer the proof of (\ref{eqn-coverlarge}) to the next section.  

Using (\ref{eqn-coversmall}) and (\ref{eqn-coverlarge}), we can bound the integral in Dudley's inequality.  We get 
\begin{equation}
L_0 \leq C_4 R\sqrt{r} K \log^{5/2} d \log^{1/2} m,
\end{equation}
where $C_4$ is some universal constant.  This proves the lemma.


\subsection{Proof of Equation (\ref{eqn-coverlarge}) (covering numbers of the nuclear-norm ball)}

Our result will follow easily from a bound on covering numbers introduced in \cite{Guedon08} (where it appears as Lemma 1):
\begin{lem}
Let $E$ be a Banach space, having modulus of convexity of power type 2 with constant $\lambda(E)$.  Let $E^*$ be the dual space, and let $T_2(E^*)$ denote its type 2 constant.  Let $B_E$ denote the unit ball in $E$.

Let $V_1,\ldots,V_m \in E^*$, such that $\norm{V_j}_{E^*} \leq K$ (for all $j$).  Define the norm on $E$, 
\begin{equation}
\norm{M}_X = \max_{j=1,\ldots,m} |(V_j,M)|.
\end{equation}
Then, for any $\vareps>0$, 
\begin{equation}
\vareps \log^{1/2} N(B_E,\norm{\cdot}_X,\vareps) \leq C_2 \lambda(E)^2 T_2(E^*) K \log^{1/2} m,
\end{equation}
where $C_2$ is some universal constant.
\end{lem}
The proof uses entropy duality to reduce the problem to bounding the ``dual'' covering number.  The basic idea is as follows.  Let $\ell_p^m$ denote the complex vector space $\CC^m$ with the $\ell_p$ norm.  Consider the map $S:\; \ell_1^m \rightarrow E^*$ that takes the $j$'th coordinate vector to $V_j$.  Let $N(S)$ denote the number of balls in $E^*$ needed to cover the image (under the map $S$) of the unit ball in $\ell_1^m$.  We can bound $N(S)$ using Maurey's empirical method.  Also define the dual map $S^*:\; E \rightarrow \ell_\infty^m$, and the associated dual covering number $N(S^*)$.  Then $N(B_E,\norm{\cdot}_X,\vareps)$ is related to $N(S^*)$.  Finally, $N(S)$ and $N(S^*)$ are related via entropy duality inequalities.  See \cite{Guedon08} for details.

We will apply this lemma as follows, using the same approach as \cite{Aubrun09}.  Let $S_p$ denote the Banach space consisting of all matrices in $\CC^{d\times d}$ with the Schatten $p$-norm.  Intuitively, we want to set $E = S_1$ and $E^* = S_\infty$, but this won't work because $\lambda(S_1)$ is infinite.  Instead, we let $E = S_p$, $p = (\log d)/(\log d-1)$, and $E^* = S_q$, $q = \log d$.  Note that $\norm{M}_p \leq \norm{M}_*$, hence $B_1 \subseteq B_p$ and 
\begin{equation}
\vareps \log^{1/2} N(B_1,\norm{\cdot}_X,\vareps)
 \leq \vareps \log^{1/2} N(B_p,\norm{\cdot}_X,\vareps).
\end{equation}
Also, we have $\lambda(E) \leq 1/\sqrt{p-1} = \sqrt{\log d-1}$ and $T_2(E^*) \leq \lambda(E) \leq \sqrt{\log d-1}$ (see the Appendix in \cite{Aubrun09}).  Note that $\norm{M}_q \leq e\norm{M}$, thus we have $\norm{V_j}_q \leq eK$ (for all $j$).  Then, using the lemma, we have 
\begin{equation}
\vareps \log^{1/2} N(B_p,\norm{\cdot}_X,\vareps) \leq C_2 \log^{3/2} d\; (eK) \log^{1/2} m,
\end{equation}
which proves the claim.


\section{Outlook}

We have showed that random Pauli measurements obey the restricted isometry property (RIP), which implies strong error bounds for low-rank matrix recovery.  The key technical tool was a bound on covering numbers of the nuclear norm ball, due to Gu\'edon et al \cite{Guedon08}.  

An interesting question is whether this method can be applied to other problems, such as matrix completion, or constructing embeddings of low-dimensional manifolds into linear spaces with slightly higher dimension.  For matrix completion, one can compare with the work of Negahban and Wainwright \cite{NW10}, where the sampling operator satisfies restricted strong convexity (RSC) over a certain set of ``non-spiky'' low-rank matrices.  For manifold embeddings, one could try to generalize the results of \cite{KW11}, which use the sparse-vector RIP to construct Johnson-Lindenstrauss metric embeddings.

There are also many questions pertaining to low-rank quantum state tomography.  For example, how does the matrix Lasso compare to the traditional approach using maximum likelihood estimation?  Also, there are several variations on the basic tomography problem, and alternative notions of sparsity (e.g., elementwise sparsity in a known basis) \cite{Shabani10}, which have not been fully explored.


\textbf{Acknowledgements:}  
Thanks to David Gross, Yaniv Plan, Emmanuel Cand\`es, Stephen Jordan, and the anonymous reviewers, for helpful suggestions.  Parts of this work were done at the University of California, Berkeley, and supported by NIST grant number 60NANB10D262.
This paper is a contribution of the National Institute of Standards and Technology, and is not subject to U.S. copyright.

\bibliography{mybib}{}
\bibliographystyle{unsrt}


\newpage

\appendix

\section*{Universal low-rank matrix recovery from Pauli measurements:}

\section*{Supplementary material}


\section{Proof of the RIP for Pauli Measurements}
\label{app3}

\subsection{Overview}

We now prove Theorem \ref{thm-rip}.  In this section we give an overview; proofs of the technical claims are deferred to later sections.  The general approach involving Dudley's entropy bound is similar to \cite{RV06}, while the technical part of the proof (bounding certain covering numbers) uses ideas from \cite{Guedon08}.

Recall the definition of the restricted isometry property, with constant $0 \leq \delta < 1$.  Let 
\begin{equation}
U = \set{X \in \CC^{d\times d} \;|\; \norm{X}_* \leq \sqrt{r}\norm{X}_F}.  
\end{equation}
Let us define 
\begin{equation}
U_2 = \set{X\in\CC^{d\times d} \;|\; \norm{X}_F \leq 1, \; \norm{X}_* \leq \sqrt{r}\norm{X}_F},
\end{equation}
\begin{equation}
\vareps_r(\calA) = \sup_{X\in U_2} |(X, (\calA^*\calA-\calI) X)|.
\end{equation}
Also, define $\vareps = 2\delta - \delta^2$.  We claim that, to show RIP, it suffices to show $\vareps_r(\calA) < \vareps$.  To see this, note that the RIP condition is equivalent to the statement 
\begin{equation}
\text{for all $X\in U$}, \quad 
(1-\delta)^2 (X, X) \leq (X, \calA^*\calA X)
 \leq (1+\delta)^2 (X, X), 
\end{equation}
which is equivalent to 
\begin{equation}
\text{for all $X\in U$}, \quad 
(-2\delta+\delta^2) (X, X) \leq (X, (\calA^*\calA-\calI) X)
 \leq (2\delta+\delta^2) (X, X), 
\end{equation}
which is implied by 
\begin{equation}
\text{for all $X\in U_2$}, \quad 
|(X, (\calA^*\calA-\calI) X)| \leq \min\set{2\delta+\delta^2, 2\delta-\delta^2} = 2\delta-\delta^2.
\end{equation}
Thus our goal is to show $\vareps_r(\calA) < \vareps$.  (Note that for $\delta$ in the range $[0,1]$, we have that $\vareps \geq \delta$.)

Let $\calB$ be the set of all self-adjoint linear operators from $\CC^{d\times d}$ to $\CC^{d\times d}$, and define the following norm on $\calB$:
\begin{equation}
\label{eqn-rnorm}
\norm{\calM}_{(r)} = \sup_{X\in U_2} |(X,\calM X)|.
\end{equation}
Suppose that $r \geq 2$ (this will suffice for our purposes, since RIP with $r=2$ implies RIP with $r=1$).  We claim that $\norm{\cdot}_{(r)}$ is a norm, and that $\calB$ is a Banach space with respect to this norm.  

To show these claims, we will consider the Frobenius norm $\norm{\cdot}_F$ on $\calB$, which is defined by viewing each element of $\calB$ as a ``matrix'' acting on ``vectors'' that are elements of $\CC^{d\times d}$.  Then we will bound $\norm{\cdot}_{(r)}$ in terms of $\norm{\cdot}_F$.  More precisely, let $\vec{e}_a$ ($a \in \set{0,1,\ldots,d-1}$) be the standard basis vectors in $\CC^d$, and let $E_{ab} = \vec{e}_a \vec{e}_b^*$ ($a,b \in \set{0,1,\ldots,d-1}$) be the standard basis vectors in $\CC^{d\times d}$.  Then the Frobenius norm on $\calB$ can be written as 
\begin{equation}
\norm{\calM}_F = \Bigl( \sum_{abcd} \Bigl| \Bra{E_{cd}}\calM\Ket{E_{ab}} \Bigr|^2 \Bigr)^{1/2}.
\end{equation}

We claim that, for all $\calM \in \calB$, 
\begin{equation} \label{eqn-rnorm-geq-fnorm}
\norm{\calM}_{(r)} \geq \frac{1}{3\sqrt{2}d^2} \norm{\calM}_F.  
\end{equation}
To see this, suppose that $\norm{\calM}_F \geq \mu$; then there must exist $a,b,c,d \in \set{0,1,\ldots,d-1}$ such that $\bigl| (E_{cd}| \calM |E_{ab}) \bigr| \geq \frac{1}{d^2} \mu$.  If $E_{ab} = E_{cd}$, then we have $\norm{\calM}_{(r)} \geq \frac{1}{d^2}\mu$.  Otherwise, we have $(E_{ab}|E_{cd}) = 0$.  Now at least one of the following must be true:  
\begin{equation}
\bigl| \text{Re} \Bra{E_{cd}}\calM\Ket{E_{ab}} \bigr| \geq \tfrac{1}{\sqrt{2} d^2} \mu \quad \text{(case 1)}, 
\end{equation}
\begin{equation}
\bigl| \text{Im} \Bra{E_{cd}}\calM\Ket{E_{ab}} \bigr| \geq \tfrac{1}{\sqrt{2} d^2} \mu \quad \text{(case 2)}.
\end{equation}
In case 1, let $X = \frac{1}{\sqrt{2}} (E_{ab} + E_{cd})$, and write 
\begin{equation}
\text{Re} \Bra{E_{cd}}\calM\Ket{E_{ab}} = \Bra{X}\calM\Ket{X} - \tfrac{1}{2} \Bra{E_{ab}}\calM\Ket{E_{ab}} - \tfrac{1}{2} \Bra{E_{cd}}\calM\Ket{E_{cd}}.
\end{equation}
One of the three terms on the right hand side must have absolute value at least $\tfrac{1}{3\sqrt{2} d^2} \mu$.  Since $X, E_{ab}, E_{cd}$ are in $U_2$, it follows that $\norm{\calM}_{(r)} \geq \tfrac{1}{3\sqrt{2} d^2} \mu$.  In case 2, let $X = \frac{1}{\sqrt{2}} (E_{ab} + iE_{cd})$, and write 
\begin{equation}
\text{Im} \Bra{E_{cd}}\calM\Ket{E_{ab}} = i\Bra{X}\calM\Ket{X} - \tfrac{1}{2} i\Bra{E_{ab}}\calM\Ket{E_{ab}} - \tfrac{1}{2} i\Bra{E_{cd}}\calM\Ket{E_{cd}}.
\end{equation}
By a similar argument, we get that $\norm{\calM}_{(r)} \geq \tfrac{1}{3\sqrt{2} d^2} \mu$.  This shows (\ref{eqn-rnorm-geq-fnorm}).

In addition, it is straightforward to see that 
\begin{equation} \label{eqn-rnorm-leq-fnorm}
\norm{\calM}_{(r)} \leq \sup_{X \::\: \norm{X}_F \leq 1} |(X, \calM X)|
 \leq \norm{\calM}_{\text{op}} \leq \norm{\calM}_F.
\end{equation}

Finally, using (\ref{eqn-rnorm-geq-fnorm}) and (\ref{eqn-rnorm-leq-fnorm}), we see that $\norm{\cdot}_{(r)}$ is a norm, and $\calB$ is a Banach space with respect to $\norm{\cdot}_{(r)}$.  (This follows since these same properties already hold for $\norm{\cdot}_F$.)  In particular, $\norm{\cdot}_{(r)}$ is nondegenerate ($\norm{\calM}_{(r)} = 0$ implies $\calM = 0$), and $\calB$ is complete with respect to $\norm{\cdot}_{(r)}$.

Returning to our main proof, we can now write $\vareps_r(\calA) = \norm{\calA^*\calA-\calI}_{(r)}$.  The strategy of the proof will be to first bound $\EE \vareps_r(\calA)$, then show that $\vareps_r(\calA)$ is concentrated around its mean.

We claim that 
\begin{equation}
\EE \vareps_r(\calA) \leq 2\EE \Norm{\sum_{j=1}^m \vareps_j \Ket{S_j}\Bra{S_j} \tfrac{d^2}{m}}_{(r)},
\end{equation}
where the $\vareps_j$ are Rademacher (iid $\pm 1$) random variables.  Here the round ket notation $\Ket{S_j}$ means we view the matrix $S_j$ as an element of the vector space $\CC^{d^2}$ with Hilbert-Schmidt inner product; the round bra $\Bra{S_j}$ denotes the adjoint element in the (dual) vector space.  The above bound follows from a standard symmetrization argument:  write $\calA^*\calA-\calI = \sum_{j=1}^m \calX_j$ where $\calX_j = \Ket{S_j}\Bra{S_j}\frac{d^2}{m} - \frac{\calI}{m}$, then let $\calX'_j$ be independent copies of the random variables $\calX_j$, and use equation (2.5) and Lemma 6.3 in \cite{LT91} to write: 
\begin{equation}
\begin{split}
\EE\vareps_r(\calA) &= \EE\Norm{\sum_j \calX_j}_{(r)} \\
 &\leq \EE\Norm{\sum_j (\calX_j-\calX'_j)}_{(r)} = \EE\norm{\sum_j \vareps_j(\calX_j-\calX'_j)}_{(r)} \\
 &= \EE\Norm{\sum_j \vareps_j \Bigl( \Ket{S_j}\Bra{S_j} - \Ket{S'_j}\Bra{S'_j} \Bigr) \tfrac{d^2}{m}}_{(r)} \\
 &\leq 2\EE\Norm{\sum_j \vareps_j \Ket{S_j}\Bra{S_j} \tfrac{d^2}{m}}_{(r)}.
\end{split}
\end{equation}

Now we use the following lemma, which we will prove later.  This bounds the expected magnitude in $(r)$-norm of a Rademacher sum of a fixed collection of operators $V_1,\ldots,V_m$ that have small operator norm.
\begin{lem}
(restatement of Lemma \ref{lem-key})
Let $m\leq d^2$.  Fix some $V_1,\ldots,V_m \in \CC^{d\times d}$ that have uniformly bounded operator norm, $\norm{V_i} \leq K$ (for all $i$).  Let $\vareps_1,\ldots,\vareps_m$ be iid uniform $\pm 1$ random variables.  Then 
\begin{equation}
\EE_\vareps \Norm{\sum_{i=1}^m \vareps_i\Ket{V_i}\Bra{V_i}}_{(r)}
 \leq C_5\cdot \Norm{\sum_{i=1}^m \Ket{V_i}\Bra{V_i}}_{(r)}^{1/2}, 
\end{equation}
where $C_5 = \sqrt{r}\cdot C_4 K \log^{5/2} d \log^{1/2} m$ and $C_4$ is some universal constant.
\end{lem}

We apply the lemma as follows.  Let $\Omega = \set{S_1,\ldots,S_m}$ be the multiset of all the measurement operators that appear in the sampling operator $\calA$.  Then we have 
\begin{equation}
\EE\vareps_r(\calA) \leq 2\EE_\Omega \EE_\vareps \Norm{\sum_{J\in\Omega} \vareps_J \sqrt{d}\Ket{J}\Bra{J}\sqrt{d}}_{(r)} \cdot \tfrac{d}{m}.
\end{equation}
Using the lemma on the set of operators $\sqrt{d}J$ ($J\in\Omega$), we get 
\begin{equation}
\begin{split}
\EE\vareps_r(\calA)
 &\leq 2\EE_\Omega C_5\cdot \Norm{\sum_{J\in\Omega} \sqrt{d}\Ket{J}\Bra{J}\sqrt{d}}_{(r)}^{1/2} \cdot \tfrac{d}{m} \\
 &\leq 2 \Bigl( \EE_\Omega \Norm{\sum_{J\in\Omega} \sqrt{d}\Ket{J}\Bra{J}\sqrt{d}}_{(r)} \Bigr)^{1/2} \cdot C_5 \cdot \tfrac{d}{m} \\
 &= 2 \Bigl( \EE \norm{\calA^*\calA}_{(r)} \Bigr)^{1/2} \cdot C_5 \cdot \sqrt{\tfrac{d}{m}} \\
 &\leq 2 (\EE\vareps_r(\calA) + 1)^{1/2} \cdot C_5 \cdot \sqrt{\tfrac{d}{m}}, 
\end{split}
\end{equation}
where $C_5 = \sqrt{r}\cdot C_4 K \log^3 d$.  

To make the notation more concise, define $E_0 = \EE\vareps_r(\calA)$ and $C_0 = 2C_5\sqrt{\tfrac{d}{m}}$.  Then, squaring both sides and rearranging, we have 
\begin{equation}
E_0^2 - C_0^2E_0 - C_0^2 \leq 0.
\end{equation}
This quadratic equation has two roots, which are given by $\alpha_\pm = \tfrac{1}{2} (C_0^2 \pm C_0\sqrt{C_0^2+4})$, and we know that $E_0$ is bounded by 
\begin{equation}
\alpha_- \leq 0 \leq E_0 \leq \alpha_+.  
\end{equation}
Also, we can simplify the bound by writing $\alpha_+ \leq \tfrac{1}{2} (C_0^2 + C_0(C_0+2)) = C_0^2 + C_0$.

Now we use the fact that $m$ is large.  Let $\lambda \geq 1$ (we will choose a precise value for $\lambda$ later).  Assume that 
\begin{equation}
m \geq \lambda d (2C_5)^2 = \lambda \cdot 4C_4^2 \cdot dr \cdot K^2 \log^6 d.  
\end{equation}
Then $C_0 \leq 1/\sqrt{\lambda}$, and $\alpha_+ \leq \tfrac{1}{\lambda} + \tfrac{1}{\sqrt{\lambda}}$, and we have the desired result:
\begin{equation}
\EE\vareps_r(\calA) \leq \tfrac{1}{\lambda} + \tfrac{1}{\sqrt{\lambda}}.
\end{equation}

It remains to show that $\vareps_r(\calA)$ is concentrated around its expectation.  We will use a concentration inequality from \cite{LT91} for sums of independent symmetric random variables that take values in some Banach space.  Define $\calX = \sum_{j=1}^m \calX_j$ where $\calX_j = \tfrac{d^2}{m} \Ket{S_j}\Bra{S_j} - \tfrac{\calI}{m}$; then we have $\calA^*\calA - \calI = \calX$ and $\vareps_r(\calA) = \norm{\calX}_{(r)}$.  

We showed above that $\EE \norm{\calX}_{(r)} \leq \tfrac{1}{\lambda} + \tfrac{1}{\sqrt{\lambda}}$.  In addition, we can bound each $\calX_j$ as follows, using the fact that, for $X\in U_2$, $|(S_j,X)| \leq \norm{S_j} \norm{X}_* \leq (K/\sqrt{d}) \sqrt{r}\norm{X}_F \leq (K/\sqrt{d}) \sqrt{r}$.
\begin{equation}
\norm{\calX_j}_{(r)}
 = \sup_{X\in U_2} \Bigl| \tfrac{d^2}{m} |(S_j,X)|^2 - \tfrac{1}{m} \norm{X}_F^2 \Bigr|
 \leq \frac{drK^2 + 1}{m} \leq \frac{1}{\lambda\cdot 4C_4^2}.
\end{equation}

We use a standard symmetrization argument:  let $\calX'_j$ denote an independent copy of $\calX_j$, and define $\calY_j = \calX_j - \calX'_j$, which is symmetric ($-\calY_j$ has the same distribution as $\calY_j$).  Also define $\calY = \sum_{j=1}^m \calY_j = \calX-\calX'$.  Using the triangle inequality, we have 
\begin{equation}
\label{eqn-EY}
\EE \norm{\calY}_{(r)} \leq 2\EE \norm{\calX}_{(r)} \leq 2(\tfrac{1}{\lambda} + \tfrac{1}{\sqrt{\lambda}}),
\end{equation}
\begin{equation}
\label{eqn-Yj}
\norm{\calY_j}_{(r)} \leq 2\norm{\calX_j}_{(r)} \leq \frac{1}{\lambda\cdot 2C_4^2}.
\end{equation}
Using equation (6.1) in \cite{LT91}, we have, for any $u\geq 0$,
\begin{equation}
\label{eqn-Xtail}
\Pr\Bigl[ \norm{\calX}_{(r)} > 2(\tfrac{1}{\lambda} + \tfrac{1}{\sqrt{\lambda}}) + u \Bigr]
 \leq \Pr\Bigl[ \norm{\calX}_{(r)} > 2\EE \norm{\calX}_{(r)} + u \Bigr]
 \leq 2\Pr\Bigl[ \norm{\calY}_{(r)} > u \Bigr].
\end{equation}

We will use the following concentration inequality of Ledoux and Talagrand \cite{LT91}.  This is a special case of Theorem 6.17 in \cite{LT91}, where we set $s=R\ell$ and use equation (6.19) in \cite{LT91}.  This is the same bound used in \cite{RV06}.
\begin{thm}
Let $\calY_1,\ldots,\calY_m$ be independent symmetric random variables taking values in some Banach space.  Assume that $\norm{\calY_j} \leq R$ for all $j$.  Let $\calY = \sum_{j=1}^m \calY_j$.  Then, for any integers $\ell \geq q$, and any $t>0$, 
\begin{equation}
\Pr\Bigl[ \norm{\calY} \geq 8q\EE\norm{\calY} + 2R\ell + t\EE\norm{\calY} \Bigr]
 \leq (C_7/q)^\ell + 2\exp(-t^2/256q),
\end{equation}
where $C_7$ is some universal constant.
\end{thm}
Now set $q = \lceil eC_7 \rceil$.  Introduce a new parameter $s \geq \sqrt{q} + 1$, and set $\ell = \lfloor s^2 \rfloor$ and $t = s$.  We get that the failure probability is exponentially small in $s$:
\begin{equation}
\Pr\Bigl[ \norm{\calY}_{(r)} \geq (8q + s) \EE\norm{\calY}_{(r)} + 2R s^2 \Bigr]
 \leq e^{-s^2+1} + 2e^{-s^2/256q}.
\end{equation}
Then, using (\ref{eqn-EY}), (\ref{eqn-Yj}) and (\ref{eqn-Xtail}), we get 
\begin{equation}
\Pr\Bigl[ \norm{\calX}_{(r)} \geq (1 + 8q + s) \cdot 2(\tfrac{1}{\lambda}+\tfrac{1}{\sqrt{\lambda}}) + \tfrac{1}{\lambda C_4^2} s^2 \Bigr]
 \leq 2[e^{-s^2+1} + 2e^{-s^2/256q}].
\end{equation}
Now let $\lambda \geq (1+8q)^2 \cdot \frac{256}{\vareps^2}$ (note that $\lambda \geq 1$, as required).  Then set $s = \frac{\vareps\sqrt{\lambda}}{16}$ (note that $s \geq 1+8q \geq \sqrt{q} + 1$, as required).  Then we can write 
\begin{equation}
\begin{split}
(1+8q+s) \cdot 2(\tfrac{1}{\lambda}+\tfrac{1}{\sqrt{\lambda}}) + \tfrac{1}{\lambda C_4^2} s^2
 \leq \tfrac{8s}{\sqrt{\lambda}} + \tfrac{s^2}{C_4^2\lambda}
 = \tfrac{\vareps}{2} + \tfrac{\vareps^2}{256C_4^2} \leq \vareps.
\end{split}
\end{equation}
Plugging into the previous inequality, we have 
\begin{equation}
\Pr[ \norm{\calX}_{(r)} \geq \vareps ] \leq e^{-\Omega(s^2)} = e^{-\Omega(\vareps^2 \lambda)}.
\end{equation}
Therefore, we have $\vareps_r(\calA) \leq \vareps$, with a failure probability that decreases exponentially in $\lambda$.  This completes the proof.


\subsection{Proof of Lemma \ref{lem-key} (bounding a Rademacher sum in $(r)$-norm)}

Let $L_0 = \EE_\vareps \norm{\sum_{i=1}^m \vareps_i\Ket{V_i}\Bra{V_i}}_{(r)}$; this is the quantity we want to bound.  We can upper-bound it by replacing the $\pm 1$ random variables $\vareps_1,\ldots,\vareps_m$ with iid $N(0,1)$ Gaussian random variables $g_1,\ldots,g_m$ (see Lemma 4.5 and equation (4.8) in \cite{LT91}); then we get 
\begin{equation}
L_0 \leq \EE_g \Norm{\sqrt{\tfrac{\pi}{2}} \sum_{i=1}^m g_i\Ket{V_i}\Bra{V_i}}_{(r)}.
\end{equation}
Using the definition of the norm $\norm{\cdot}_{(r)}$ (equation (\ref{eqn-rnorm})), we have 
\begin{equation}
L_0 \leq \EE_g \sup_{X\in U_2} \sqrt{\tfrac{\pi}{2}} |G(X)|, \quad
G(X) = \sum_{i=1}^m g_i |(V_i,X)|^2.
\end{equation}
The random variables $G(X)$ (indexed by $X\in U_2$) form a Gaussian process, and $L_0$ is upper-bounded by the expected supremum of this process.  In particular, using the fact that $G(0)=0$ and $G(\cdot)$ is symmetric (see \cite{LT91}, pp.298), we have 
\begin{equation}
\begin{split}
L_0 &\leq \sqrt{\tfrac{\pi}{2}} \EE_g \sup_{X\in U_2} |G(X)-G(0)|
 \leq \sqrt{\tfrac{\pi}{2}} \EE_g \sup_{X,Y\in U_2} |G(X)-G(Y)| \\
 &= \sqrt{\tfrac{\pi}{2}} \EE_g \sup_{X,Y\in U_2} G(X)-G(Y)
 = \sqrt{2\pi} \EE_g \sup_{X\in U_2} G(X).
\end{split}
\end{equation}
Using Dudley's inequality (Theorem 11.17 in \cite{LT91}), we have 
\begin{equation}
\label{eqn-dudley}
L_0 \leq 24 \sqrt{2\pi} \int_0^\infty \log^{1/2} N(U_2,d_G,\vareps) d\vareps, 
\end{equation}
where $N(U_2,d_G,\vareps)$ is a covering number (the number of balls in $\CC^{d\times d}$ of radius $\vareps$ in the metric $d_G$ that are needed to cover the set $U_2$), and the metric $d_G$ is given by 
\begin{equation}
d_G(X,Y) = \Bigl( \EE[(G(X)-G(Y))^2] \Bigr)^{1/2}.
\end{equation}

We can simplify the metric $d_G$, using the fact that $\EE[g_ig_j] = 1$ when $i=j$ and 0 otherwise:
\begin{equation}
\begin{split}
d_G(X,Y) &= \Bigl( \EE\Bigl[ \Bigl( \sum_{i=1}^m g_i (|(V_i,X)|^2-|(V_i,Y)|^2) \Bigr)^2 \Bigr] \Bigr)^{1/2} \\
 &= \Bigl( \sum_{i=1}^m \Bigl( |(V_i,X)|^2-|(V_i,Y)|^2 \Bigr)^2 \Bigr)^{1/2}
\end{split}
\end{equation}
Define a new norm (actually a semi-norm) $\norm{\cdot}_X$ on $\CC^{d\times d}$, as follows:
\begin{equation}
\norm{M}_X = \max_{i=1,\ldots,m} |(V_i,M)|.
\end{equation}
Note that 
\footnote{Note that, for any complex numbers $a$ and $b$, $|a|^2 - |b|^2
 = \tfrac{1}{2} (\bar{a}+\bar{b}) (a-b) + \tfrac{1}{2} (a+b) (\bar{a}-\bar{b})
 \leq |a+b|\cdot |a-b|$.}
\begin{equation}
\begin{split}
\Bigl| |(V_i,X)|^2-|(V_i,Y)|^2 \Bigr|
 &\leq \Bigl( |(V_i,X)|+|(V_i,Y)| \Bigr) \cdot |(V_i,X)-(V_i,Y)| \\
 &\leq \Bigl( |(V_i,X)|+|(V_i,Y)| \Bigr) \cdot \norm{X-Y}_X.
\end{split}
\end{equation}
This lets us give a simpler upper bound on the metric $d_G$:
\begin{equation}
\begin{split}
d_G(X,Y)
 &\leq \Bigl( \sum_{i=1}^m \Bigl( |(V_i,X)|+|(V_i,Y)| \Bigr)^2 \cdot \norm{X-Y}_X^2 \Bigr)^{1/2} \\
 &\leq \Bigl[ \Bigl( \sum_{i=1}^m |(V_i,X)|^2 \Bigr)^{1/2} + \Bigl( \sum_{i=1}^m |(V_i,Y)|^2 \Bigr)^{1/2} \Bigr] \cdot \norm{X-Y}_X \\
 &\leq 2\sup_{X\in U_2} \Bigl( \sum_{i=1}^m |(V_i,X)|^2 \Bigr)^{1/2} \cdot \norm{X-Y}_X \\
 &= 2\Norm{\sum_{i=1}^m \Ket{V_i}\Bra{V_i}}_{(r)}^{1/2} \cdot \norm{X-Y}_X.
\end{split}
\end{equation}
Note that the last step holds for all $X,Y\in U_2$.  To simplify the notation, let $R = \norm{\sum_{i=1}^m \Ket{V_i}\Bra{V_i}}_{(r)}^{1/2}$, then we have $d_G(X,Y) \leq 2R\norm{X-Y}_X$.

This lets us upper-bound the covering numbers in $d_G$ with covering numbers in $\norm{\cdot}_X$:
\begin{equation}
N(U_2, d_G, \vareps) \leq N(U_2, \norm{\cdot}_X, \tfrac{\vareps}{2R}) = N(\tfrac{1}{\sqrt{r}}U_2, \norm{\cdot}_X, \tfrac{\vareps}{2R\sqrt{r}}).
\end{equation}
Plugging into (\ref{eqn-dudley}) and changing variables, we get 
\begin{equation}
\label{eqn-coverint}
L_0 \leq 48 \sqrt{2\pi} R\sqrt{r} \int_0^\infty \log^{1/2} N(\tfrac{1}{\sqrt{r}}U_2, \norm{\cdot}_X, \vareps) d\vareps.
\end{equation}

We will now bound these covering numbers.  First, we introduce some notation:  let $\norm{\cdot}_p$ denote the Schatten $p$-norm on $\CC^{d\times d}$, and let $B_p$ be the unit ball in this norm.  Also, let $B_X$ be the unit ball in the $\norm{\cdot}_X$ norm.  

Observe that 
\begin{equation}
\tfrac{1}{\sqrt{r}}U_2 \subseteq B_1 \subseteq K\cdot B_X.
\end{equation}
(The second inclusion follows because $\norm{M}_X \leq \max_{i=1,\ldots,m} \norm{V_i} \norm{M}_* \leq K \norm{M}_*$.)  This gives a simple bound on the covering numbers:
\begin{equation}
N(\tfrac{1}{\sqrt{r}}U_2, \norm{\cdot}_X, \vareps) \leq N(B_1, \norm{\cdot}_X, \vareps)
 \leq N(K\cdot B_X, \norm{\cdot}_X, \vareps).
\end{equation}
This equals 1 when $\vareps\geq K$.  So, in equation (\ref{eqn-coverint}), we can restrict the integral to the interval $[0,K]$.

When $\vareps$ is small, we will use the following simple bound (equation (5.7) in \cite{Pisier}):  (this is equation (\ref{eqn-coversmall}))
\begin{equation}
N(K\cdot B_X, \norm{\cdot}_X, \vareps) \leq (1+\tfrac{2K}{\vareps})^{2d^2}.
\end{equation}
When $\vareps$ is large, we will use a more sophisticated bound based on Maurey's empirical method and entropy duality, which is due to \cite{Guedon08} (see also \cite{Aubrun09}):  (this is equation (\ref{eqn-coverlarge}))
\begin{equation}
N(B_1, \norm{\cdot}_X, \vareps) \leq \exp(\tfrac{C_1^2 K^2}{\vareps^2} \log^3 d \log m), \quad \text{for some constant $C_1$}.
\end{equation}
We defer the proof of (\ref{eqn-coverlarge}) to the next section.  Here, we proceed to bound the integral in (\ref{eqn-coverint}).

Let $A = K/d$.  For the integral over $[0,A]$, we write 
\begin{equation}
\begin{split}
L_1 &:= \int_0^A \log^{1/2} N(\tfrac{1}{\sqrt{r}}U_2, \norm{\cdot}_X, \vareps) d\vareps
 \leq \int_0^A \sqrt{2}d \log^{1/2} (1+\tfrac{2K}{\vareps}) d\vareps \\
 &\leq \sqrt{2}d \int_0^A \bigl( 1+\log(1+\tfrac{2K}{\vareps}) \bigr) d\vareps
 = \sqrt{2}d\cdot A + \sqrt{2}d\cdot L'_1
\end{split}
\end{equation}
where 
\begin{equation}
\begin{split}
L'_1 &:= \int_0^A \log(1+\tfrac{2K}{\vareps}) d\vareps
 = \int_{1/A}^\infty \log(1+2Ky) \tfrac{dy}{y^2} \\
 &\leq \int_{1/A}^\infty \log((A+2K)y) \tfrac{dy}{y^2}
 = \int_{1/A}^\infty \log(A+2K) \tfrac{dy}{y^2} + \int_{1/A}^\infty \log y \tfrac{dy}{y^2}.
\end{split}
\end{equation}
Integrating by parts, we get 
\begin{equation}
L'_1 \leq A\log(A+2K) + A\log\tfrac{1}{A} + A = A\log(1+\tfrac{2K}{A}) + A,
\end{equation}
and substituting back in,
\begin{equation}
L_1 \leq \sqrt{2}dA (2+\log(1+\tfrac{2K}{A})) = \sqrt{2}K (2+\log(1+2d)).
\end{equation}

For the integral over $[A,K]$, we write 
\begin{equation}
\begin{split}
L_2 &:= \int_A^K \log^{1/2} N(\tfrac{1}{\sqrt{r}}U_2, \norm{\cdot}_X, \vareps) d\vareps
 \leq \int_A^K \tfrac{C_1 K}{\vareps} \log^{3/2} d \log^{1/2} m\; d\vareps \\
 &= C_1 K \log^{3/2} d \log^{1/2} m \log \tfrac{K}{A}
 = C_1 K \log^{5/2} d \log^{1/2} m.
\end{split}
\end{equation}

Finally, substituting into (\ref{eqn-coverint}), we get 
\begin{equation}
L_0 \leq 48 \sqrt{2\pi} R\sqrt{r} (L_1+L_2)
 \leq C_4 R\sqrt{r} K \log^{5/2} d \log^{1/2} m,
\end{equation}
where $C_4$ is some universal constant.  This proves the lemma.


\subsection{Proof of Equation (\ref{eqn-coverlarge}) (covering numbers of the nuclear-norm ball)}

Our result will follow easily from a bound on covering numbers introduced in \cite{Guedon08} (where it appears as Lemma 1):
\begin{lem}
Let $E$ be a Banach space, having modulus of convexity of power type 2 with constant $\lambda(E)$.  Let $E^*$ be the dual space, and let $T_2(E^*)$ denote its type 2 constant.  Let $B_E$ denote the unit ball in $E$.

Let $V_1,\ldots,V_m \in E^*$, such that $\norm{V_j}_{E^*} \leq K$ (for all $j$).  Define the norm on $E$, 
\begin{equation}
\norm{M}_X = \max_{j=1,\ldots,m} |(V_j,M)|.
\end{equation}
Then, for any $\vareps>0$, 
\begin{equation}
\vareps \log^{1/2} N(B_E,\norm{\cdot}_X,\vareps) \leq C_2 \lambda(E)^2 T_2(E^*) K \log^{1/2} m,
\end{equation}
where $C_2$ is some universal constant.
\end{lem}
The proof uses entropy duality to reduce the problem to bounding the ``dual'' covering number.  The basic idea is as follows.  Let $\ell_p^m$ denote the complex vector space $\CC^m$ with the $\ell_p$ norm.  Consider the map $S:\; \ell_1^m \rightarrow E^*$ that takes the $j$'th coordinate vector to $V_j$.  Let $N(S)$ denote the number of balls in $E^*$ needed to cover the image (under the map $S$) of the unit ball in $\ell_1^m$.  We can bound $N(S)$ using Maurey's empirical method.  Also define the dual map $S^*:\; E \rightarrow \ell_\infty^m$, and the associated dual covering number $N(S^*)$.  Then $N(B_E,\norm{\cdot}_X,\vareps)$ is related to $N(S^*)$.  Finally, $N(S)$ and $N(S^*)$ are related via entropy duality inequalities.  See \cite{Guedon08} for details.

We will apply this lemma as follows, using the same approach as \cite{Aubrun09}.  Let $S_p$ denote the Banach space consisting of all matrices in $\CC^{d\times d}$ with the Schatten $p$-norm.  Intuitively, we want to set $E = S_1$ and $E^* = S_\infty$, but this won't work because $\lambda(S_1)$ is infinite.  Instead, we let $E = S_p$, $p = (\log d)/(\log d-1)$, and $E^* = S_q$, $q = \log d$.  Note that $\norm{M}_p \leq \norm{M}_*$, hence $B_1 \subseteq B_p$ and 
\begin{equation}
\vareps \log^{1/2} N(B_1,\norm{\cdot}_X,\vareps)
 \leq \vareps \log^{1/2} N(B_p,\norm{\cdot}_X,\vareps).
\end{equation}
Also, we have $\lambda(E) \leq 1/\sqrt{p-1} = \sqrt{\log d-1}$ and $T_2(E^*) \leq \lambda(E) \leq \sqrt{\log d-1}$ (see the Appendix in \cite{Aubrun09}).  Note that $\norm{M}_q \leq e\norm{M}$, thus we have $\norm{V_j}_q \leq eK$ (for all $j$).  Then, using the lemma, we have 
\begin{equation}
\vareps \log^{1/2} N(B_p,\norm{\cdot}_X,\vareps) \leq C_2 \log^{3/2} d\; (eK) \log^{1/2} m,
\end{equation}
which proves the claim.


\section{Proof of Proposition \ref{prop-errorbound2} (recovery of a full-rank matrix)}
\label{sec4}

In this section we will sketch the proof of Proposition \ref{prop-errorbound2}.  We use the same argument as Theorem 2.8 in \cite{CP09}, adapted for Pauli (rather than Gaussian) measurements.  

A crucial ingredient is the NNQ (``nuclear norm quotient'') property of a sampling operator $\calA$, which was introduced in \cite{CP09} and is analogous to the LQ (``$\ell_1$-quotient'') property in compressed sensing \cite{Woj09}.  We say that a sampling operator $\calA:\: \CC^{d\times d} \rightarrow \CC^m$ satisfies the NNQ($\alpha$) property if 
\begin{equation}
\calA(B_1) \supseteq \alpha B_2,
\end{equation}
where $B_1$ is the unit ball of the nuclear norm in $\CC^{d\times d}$, and $B_2$ is the unit ball of the $\ell_2$ (Euclidean) norm in $\CC^m$.

It is easy to see that the Pauli sampling operator $\calA$ defined in (\ref{eqn-A}) satisfies NNQ($\alpha$) with $\alpha = \sqrt{d/m}$.  (Without loss of generality, suppose that the Pauli matrices $S_1,\ldots,S_m$ used to construct $\calA$ are all distinct.  Let $\alpha = \sqrt{d/m}$ and choose any $y \in \alpha B_2$.  Let $X = \frac{\sqrt{m}}{d} \sum_{i=1}^m y_i S_i$, so we have $\calA(X) = y$.  Observe that $\norm{X}_* \leq \sqrt{d} \norm{X}_F = \sqrt{\frac{m}{d}} \norm{y}_2 \leq 1$, as desired.)  We remark that this value of $\alpha$ is probably not optimal; if one could prove that $\calA$ satisfies NNQ($\alpha$) with larger $\alpha$, it would improve the bound in Proposition \ref{prop-errorbound2}.

We will need one more property of $\calA$.  We want the following to hold:  for any fixed matrix $M\in\CC^{d\times d}$ (which is not necessarily low-rank), almost all random choices of $\calA$ will satisfy 
\begin{equation}
\norm{\calA(M)}_2^2 \leq 1.5 \norm{M}_F^2.  
\end{equation}
(Note that this inequality is required to hold only for this one particular matrix $M$.)  In our case (random Pauli measurements), it is easy to check that $\calA$ obeys this property as well.  

The proof of Theorem 2.8 in \cite{CP09} actually implies the following more general statement, about low-rank matrix recovery when $\calA$ satisfies both RIP and NNQ:
\begin{thm}
\label{thm-errorbound3}
Let $M$ be any matrix in $\CC^{d\times d}$, and let $\sigma_1(M) \geq \sigma_2(M) \geq \cdots \geq \sigma_d(M) \geq 0$ be its singular values.  Write $M = M_r+M_c$, where $M_r$ contains the $r$ largest singular values of $M$.  Also write $M = M_0+M_e$, where $M_0$ contains only those singular values of $M$ that exceed $\lambda=16\sqrt{d}\sigma$.  

Suppose the sampling operator $\calA:\: \CC^{d\times d} \rightarrow \CC^m$ satisfies RIP (for rank-$r$ matrices in $\CC^{d\times d}$), and NNQ($\alpha$) with $\alpha = \mu\sqrt{d/m}$.  Furthermore, suppose that $\calA$ satisfies $\norm{\calA(M_c)}_2^2 \leq 1.5 \norm{M_c}_F^2$ and $\norm{\calA(M_e)}_2^2 \leq 1.5 \norm{M_e}_F^2$.  

Say we observe $y = \calA(M)+z$, where $z \sim N(0,\sigma^2 I)$.  Let $\Mhat$ be the Dantzig selector (\ref{eqn-tracemin}) with $\lambda = 16\sqrt{d}\sigma$, or the Lasso (\ref{eqn-lasso}) with $\mu = 32\sqrt{d}\sigma$.  Then, with high probability over the choice of $\calA$ and the noise $z$, 
\begin{equation}
\label{eqn-errorbound3}
\norm{\Mhat-M}_F^2 \leq C_0 \sum_{i=1}^r \min(\sigma_i^2(M), d\sigma^2) + \Bigl(C_1 + \frac{C_2 m}{\mu^2 rd}\Bigr) \sum_{i=r+1}^d \sigma_i^2(M), 
\end{equation}
where $C_0$, $C_1$ and $C_2$ are absolute constants.
\end{thm}

To prove Theorem \ref{thm-errorbound3}, one follows the proof of Theorem 2.8 in \cite{CP09}.  There is a slight modification to Lemma 3.10 in \cite{CP09}:  one gets the more general bound, 
\begin{equation}
\norm{\Mhat-M}_F \leq C_0 \lambda\sqrt{r} + \bigl(C_1 + \tfrac{C_2}{\mu} \sqrt{\tfrac{m}{rd}}\bigr) \norm{\calA(M_c)}_2 + \norm{M_c}_F.
\end{equation}

Combining Theorem \ref{thm-errorbound3} with the preceding facts gives us Proposition \ref{prop-errorbound2}.


\end{document}